# Analytical solution of the Eliashberg equations for strong-coupling superconductivity in hydrides

Tomas J. Escamilla and Chumin Wang*
*Instituto de Investigaciones en Materiales, Universidad Nacional Autónoma de México, MEXICO*
*Corresponding author. E-mail address: chumin@unam.mx
E-mail address: tomasjel@ciencias.unam.mx (T.J. Escamilla)

## Abstract

The recent discovery of room-temperature superconductivity in hydrogen-rich materials under extreme pressures has renewed interest in phonon-mediated pairing mechanisms described by the Eliashberg theory, which generalizes the BCS framework by incorporating phonon retardation effects. In this article, we present an analytical solution to the isotropic Eliashberg equations for the superconducting critical temperature, formulated within the Debye model. This solution exhibits good agreement with fully self-consistent numerical calculations across both weak- and strong-coupling regimes, correctly reproducing the exponential and square-root dependences on the electron-phonon coupling parameter. We further apply the solution to $YH_6$, benchmarking against experiment and the McMillan-Allen-Dynes formula. More broadly, across a diverse set of hydride superconductors, the predicted critical temperature shows scaling consistency with ab initio calculations and experimental data.

**Keywords**: Eliashberg theory; electron-phonon interaction; strong-coupling regime; hydride superconductors;

## I. Introduction

Superconductivity was first discovered in 1911 by H. K. Onnes, who observed the complete disappearance of electrical resistance in mercury at 4.15 K [1]. Throughout most of the twentieth century, experimental studies of superconductors required cooling with liquid helium due to their low critical temperatures [2]. A significant breakthrough occurred in 1986 with the discovery of ceramic high-temperature superconductors, which enabled the use of liquid nitrogen as a more practical coolant [3,4]. Despite this progress, achieving superconductivity at or near room temperature remained a major scientific challenge and a long-standing goal in condensed matter physics.

In 2015, the observation of superconductivity in compressed $H_3S$ at 203 K and 200 GPa marked the first realization of a hydride superconductor operating near ambient temperature [5], whose structural phase Im-3m was determined by in situ X-ray diffraction [6]. This discovery prompted a rapid expansion of research on hydrogen-rich materials under high pressure. Since then, numerous hydride superconductors have been experimentally confirmed, including $YH_{10}$ [7], $LaH_{10}$ [8], $ThH_9$ [9], $ThH_{10}$ [9], $YH_9$ [10], $YH_6$ [11], and $CaH_6$ [12]. These compounds share an elevated hydrogen content, which increases vibrational frequencies and enhances Cooper pairing strength, as first proposed by N. W. Ashcroft in 1968 [13], while its later extension to metallic hydrides as realistic candidates was developed by Ashcroft in 2004 [14].

Theoretical studies of phonon-mediated superconductivity in these materials are typically carried out within the Eliashberg framework, which incorporates phonon retardation effects beyond the BCS theory [15]. Moreover, the Eliashberg equations are commonly formulated in imaginary time using Matsubara frequencies, enabling a self-consistent determination of the superconducting gap $[\Delta(i\omega_n)]$ and critical temperature ($T_c$) from a given electron-phonon spectral function $[\alpha^2F(\omega)]$.

Starting from the linearized Eliashberg equations, W. L. McMillan [16] derived an analytical expression for $T_c$ which was later refined by P. B. Allen and R. C. Dynes through empirical corrections fitted to a broad set of known superconductors [17]. Notably, the superconducting properties and pressure conditions of $H_3S$ were successfully predicted using the McMillan-Allen-Dynes formula combined with the Eliashberg spectral function obtained from *ab initio* calculations [18]. Recently, the Einstein-boson model has frequently been employed to provide analytical forms of $\alpha^2F(\omega)$, thereby complementing the McMillan-Allen-Dynes solution to the isotropic Eliashberg equations [19-24]. It is well known that a more realistic description of phonon spectra is offered by the Debye model. However, its incorporation into analytical solutions of the isotropic Eliashberg equations introduces significant mathematical complexity. As a result, analytical approaches based on the Debye model have so far remained largely restricted to the leading-term approximation [25].

It is worth noting that the recent discovery of room-temperature superconductivity in hydrogen-rich materials has pushed the boundaries of electron-phonon coupling strength and characteristic frequency scales into regimes where the validity of the McMillan-Allen-Dynes parameters is not firmly established. In this article, we present a concise derivation of the isotropic Eliashberg equations and develop a wide-coupling-range analytical solution for the superconducting critical temperature within the Debye model. The Debye frequency is calculated from the integrated *ab initio* Eliashberg spectral function, *i.e.*, by using the electron-phonon coupling spectrum rather than the purely vibrational density of states.

This solution is first validated through self-consistent numerical calculations employing the Debye spectral function. It is then further verified by applying it to a strong-coupling hydride superconductor, $YH_6$, where the obtained $T_c$ is compared with results from the McMillan-Allen-Dynes formula and with available experimental data, including an analysis of the effects of the broadening parameter on both the spectral function and the predicted $T_c$. This verification is now extended to eight additional hydride compounds, revealing a strong correlation between the self-consistent and analytical solutions of the Eliashberg equations when the respective ab initio spectral functions are used in both approaches. Moreover, the robustness of this wide-range analytical solution, based on a unique Debye spectral function, is investigated by comparing the predicted $T_c$ as a function of the electron-phonon coupling parameter with a broader set of theoretical and experimental data. Finally, a detailed deduction of the analytical solution is presented in Appendix A.

## II. Concise derivation of isotropic Eliashberg equations

The Eliashberg formalism extends the BCS theory by explicitly incorporating retardation effects on the electron-electron via phonon interaction, and it can be started from a 2×2 matrix Green's function given by [26, 27],

$$\hat{G}(\mathbf{k},\tau) = -<\hat{T}_\tau \mathbf{\Psi_k}(\tau)\mathbf{\Psi_k^\dagger}(0)> = \begin{pmatrix} G(\mathbf{k},\tau) & F(\mathbf{k},\tau) \\ F^*(\mathbf{k},\tau) & -G(-\mathbf{k},-\tau) \end{pmatrix}, \quad (1)$$

where $\hat{T}_\tau$ is the Wick's time-ordering operator, τ is the imaginary time, $\mathbf{\Psi_k}(\tau) = \left(\hat{c}_{\mathbf{k},\uparrow}(\tau), \hat{c}^\dagger_{-\mathbf{k},\downarrow}(0)\right)^{\mathrm{T}}$ is a two-component Nambu spinor with $\hat{c}^\dagger_{\mathbf{k},\sigma}$ and $\hat{c}_{\mathbf{k},\sigma}$ denoting the creation and annihilation operators for an electron with momentum $\mathbf{k}$ and spin $\sigma \in \{\uparrow,\downarrow\}$. The normal electron propagation is described by $G(\mathbf{k},\tau) = -<\hat{T}_\tau\, \hat{c}_{\mathbf{k},\uparrow}(\tau)\, \hat{c}^\dagger_{\mathbf{k},\uparrow}(0)>$, while $F(\mathbf{k},\tau) = -<\hat{T}_\tau\, \hat{c}_{\mathbf{k},\uparrow}(\tau)\, \hat{c}_{-\mathbf{k},\downarrow}(0)>$ represents the anomalous Gor'kov Green's function associated with Cooper pair dynamics.

The Fourier transform of $\hat{G}(\mathbf{k},\tau)$ in atomic units $(k_{\mathrm{B}} = \hbar = 1)$ is

$$\hat{G}(\mathbf{k},i\omega_n) = \int_0^\beta d\tau\, e^{i\omega_n\tau}\hat{G}(\mathbf{k},\tau), \quad (2)$$

where $\beta = 1/T$ and $\omega_n = (2n+1)\pi T$ is the fermion Matsubara frequency with integer index *n*. The Dyson equation given by [28] $\hat{G}(\mathbf{k},i\omega_n) = \hat{G}_0(\mathbf{k},i\omega_n) + \hat{G}_0(\mathbf{k},i\omega_n)\hat{\Sigma}(\mathbf{k},i\omega_n)\hat{G}(\mathbf{k},i\omega_n)$ can be rewritten as

$$\hat{G}^{-1}(\mathbf{k},i\omega_n) = \hat{G}_0^{-1}(\mathbf{k},i\omega_n) - \hat{\Sigma}(\mathbf{k},i\omega_n), \quad (3)$$

where $\hat{G}_0(\mathbf{k},i\omega_n) = [i\omega_n\sigma_0 - \varepsilon(\mathbf{k})\sigma_z]^{-1}$ is the Green's function for electrons in the normal state with dispersion relation $\varepsilon(\mathbf{k})$ and $\hat{\Sigma}(\mathbf{k},i\omega_n)$ is the 2×2 self-energy matrix that can be expressed in terms of the Pauli matrices ($\sigma_x$, $\sigma_y$ and $\sigma_z$) and the 2×2 identity matrix ($\sigma_0$) as [29]

$$\hat{\Sigma}(\mathbf{k},i\omega_n) = i\omega_n[1 - Z(\mathbf{k},i\omega_n)]\sigma_0 + \chi(\mathbf{k},i\omega_n)\sigma_z + \phi(\mathbf{k},i\omega_n)\sigma_x + \tilde{\phi}(\mathbf{k},i\omega_n)\sigma_y\,. \quad (4)$$

Substituting Eq. (4) into Eq. (3), we get the Green's function,

$$\hat{G}(\mathbf{k},i\omega_n) = \frac{1}{\det(\hat{G}^{-1})}\Big[i\omega_n Z(\mathbf{k},i\omega_n)\,\sigma_0 + \big[\varepsilon(\mathbf{k}) + \chi(\mathbf{k},i\omega_n)\big]\sigma_z + \phi(\mathbf{k},i\omega_n)\sigma_x + \tilde{\phi}(\mathbf{k},i\omega_n)\sigma_y\Big], \quad (5)$$

where $\phi(\mathbf{k},i\omega_n)$ and $\tilde{\phi}(\mathbf{k},i\omega_n)$ are related to off-diagonal elements $F(\mathbf{k},\tau) = -<\hat{T}_\tau\, \hat{c}_{\mathbf{k},\uparrow}(\tau)\,\hat{c}_{-\mathbf{k},\downarrow}(0)>$ and

$$\det(\hat{G}^{-1}) = (i\omega_n Z)^2 - [\varepsilon(\mathbf{k}) + \chi]^2 - \phi^2 - \tilde{\phi}^2\,. \quad (6)$$

Performing an analytic continuation $i\omega_n \to E_{\mathbf{k}} + i\eta$ with $\eta \to 0^+$, the poles of $\hat{G}(\mathbf{k},i\omega_n)$ in Eq. (5) are obtained from $\det(\hat{G}^{-1}) = 0$, yielding the quasiparticle excitation spectrum given by [29]

$$E_{\mathbf{k}} = \sqrt{\frac{[\varepsilon(\mathbf{k}) + \chi(\mathbf{k},i\omega_n)]^2}{[Z(\mathbf{k},i\omega_n)]^2} + \frac{[\phi(\mathbf{k},i\omega_n)]^2 + [\tilde{\phi}(\mathbf{k},i\omega_n)]^2}{[Z(\mathbf{k},i\omega_n)]^2}} = \sqrt{\frac{[\varepsilon(\mathbf{k}) + \chi(\mathbf{k},i\omega_n)]^2}{[Z(\mathbf{k},i\omega_n)]^2} + |\Delta(\mathbf{k},i\omega_n)|^2}\,, \quad (7)$$

where $Z(\mathbf{k},i\omega_n)$ is the mass renormalization function, $\chi(\mathbf{k},i\omega_n)$ is the energy shift, and $\Delta(\mathbf{k},i\omega_n)$ denotes the superconducting gap in the Matsubara frequency domain, defined as

$$\Delta(\mathbf{k},i\omega_n)=\frac{\phi(\mathbf{k},i\omega_n)-i\tilde{\phi}(\mathbf{k},i\omega_n)}{Z(\mathbf{k},i\omega_n)}. \tag{8}$$

Hence, for s-wave superconductivity, we may choose a gauge such that $\tilde{\phi}=0$ and $\Delta(\mathbf{k},i\omega_n)\in\mathbb{R}$.

According to Migdal's theorem, the self-energy in Eq. (3) can be written as [30, 31]

$$\hat{\Sigma}(\mathbf{k},i\omega_n)\simeq -T\sum_{\mathbf{k}',n'}\sigma_z\hat{G}(\mathbf{k}',i\omega_{n'})\sigma_z\sum_{\nu}\left|g^{\nu}_{k,k'}\right|^2 D_\nu(\mathbf{k}-\mathbf{k}',i\omega_n-i\omega_{n'})-T\sum_{\mathbf{k}',n'}\sigma_z\hat{G}^{\mathrm{od}}(\mathbf{k}',i\omega_{n'})\sigma_z\tilde{V}(\mathbf{k}-\mathbf{k}'), \tag{9}$$

where $g^{\nu}_{k,k'}$ denote the electron-phonon interaction matrix elements, $\hat{G}^{\mathrm{od}}(\mathbf{k}',i\omega_{n'})=(\phi\sigma_x+\tilde{\phi}\sigma_y)/\det(\hat{G}^{-1})$ is the off-diagonal part of $\hat{G}(\mathbf{k}',i\omega_{n'})$ with null diagonal elements, and $\tilde{V}(\mathbf{k}-\mathbf{k}')$ is the screened Coulomb potential, while $D_\nu(\mathbf{k}-\mathbf{k}',i\omega_n-i\omega_{n'})$ is the phonon correlation function with wave vector $\mathbf{q}=\mathbf{k}-\mathbf{k}'$ and branch index $\nu$, corresponding to angular frequency $\omega_\nu(\mathbf{k}-\mathbf{k}')$ [29]. The first term in Eq. (9) accounts for the electron-phonon interaction, while the second one represents the screened Coulomb repulsion.

Since the Pauli matrices form a linearly independent basis, the Eliashberg equations are obtained by substituting Eqs. (4) and (5) into Eq. (9), yielding [26][32]

$$\begin{cases} Z(\mathbf{k},i\omega_n)=1+\dfrac{T}{\omega_n}\sum\limits_{\mathbf{k}',n'}\dfrac{\omega_{n'}Z(\mathbf{k}',i\omega_{n'})}{\det(G^{-1})}\sum\limits_{\nu}\left|g^{\nu}_{k,k'}\right|^2 D_\nu(\mathbf{k}-\mathbf{k}',i\omega_n-i\omega_{n'}) \\ \chi(\mathbf{k},i\omega_n)=-T\sum\limits_{\mathbf{k}',n'}\dfrac{\varepsilon(\mathbf{k}')+\chi(\mathbf{k}',i\omega_{n'})}{\det(G^{-1})}\sum\limits_{\nu}\left|g^{\nu}_{k,k'}\right|^2 D_\nu(\mathbf{k}-\mathbf{k}',i\omega_n-i\omega_{n'}) \\ \phi(\mathbf{k},i\omega_n)=T\sum\limits_{\mathbf{k}',n'}\dfrac{\phi(\mathbf{k}',i\omega_{n'})}{\det(G^{-1})}\left[\sum\limits_{\nu}\left|g^{\nu}_{k,k'}\right|^2 D_\nu(\mathbf{k}-\mathbf{k}',i\omega_n-i\omega_{n'})+\tilde{V}(\mathbf{k}-\mathbf{k}')\right]\end{cases}, \tag{10}$$

where the phonon correlation function can be approximated by its non-interacting form that is

$$D_\nu(\mathbf{k}-\mathbf{k}',i\omega_n-i\omega_{n'})\approx D^{(0)}_\nu(\mathbf{k}-\mathbf{k}',i\omega_n-i\omega_{n'})=\frac{2\omega_\nu(\mathbf{k}-\mathbf{k}')}{(i\omega_n-i\omega_{n'})^2-[\omega_\nu(\mathbf{k}-\mathbf{k}')]^2}. \tag{11}$$

The Eliashberg spectral function $\alpha^2F(\omega)$, averaged over the Fermi surface, is defined as [32]

$$\alpha^2F(\omega)=\frac{1}{N(0)}\sum_{\mathbf{k},\mathbf{k}'}\sum_{\nu}\left|g^{\nu}_{k,k'}\right|^2\delta\big(\varepsilon(\mathbf{k}')\big)\delta\big(\varepsilon(\mathbf{k}')\big)\delta\big(\omega-\omega_\nu(\mathbf{k}-\mathbf{k}')\big), \tag{12}$$

while the electron-phonon coupling constant is given by

$$\lambda(i\omega_n-i\omega_{n'})=\int_0^\infty\frac{2\omega\alpha^2F(\omega)}{(\omega_n-\omega_{n'})^2+\omega^2}d\omega. \tag{13}$$

Given that superconducting pairing mainly occurs within a narrow energy window around the Fermi level, the energy shift can be neglected by setting $\chi(\mathbf{k},i\omega_\mathrm{n})=0$. Using Eqs. (11)-(13), and averaging Eqs. (10) over the $\mathbf{k}$ space constricted at the Fermi energy $\varepsilon(\mathbf{k})=0$, the isotropic Eliashberg equations are obtained [29][32],

$$\begin{cases} Z(i\omega_n)=\langle Z(\mathbf{k},i\omega_n)\rangle_{\varepsilon(\mathbf{k})=0}=1+\dfrac{\pi T}{\omega_n}\sum\limits_{n'=-\infty}^{\infty}\dfrac{\omega_{n'}}{\sqrt{(\omega_{n'})^2+\Delta^2(i\omega_{n'})}}\lambda(i\omega_n-i\omega_{n'}) \\ \Delta(i\omega_n)=\dfrac{\langle\phi(\mathbf{k},i\omega_n)\rangle_{\varepsilon(\mathbf{k})=0}}{Z(i\omega_n)}=\dfrac{\pi T}{Z(i\omega_n)}\sum\limits_{n'=-\infty}^{\infty}\dfrac{\Delta(i\omega_{n'})}{\sqrt{(\omega_{n'})^2+\Delta^2(i\omega_{n'})}}\left[\lambda(i\omega_n-i\omega_{n'})-\mu^*\right]\end{cases}. \tag{14}$$

where $\mu^* = \mu/[1+\mu\ln(\varepsilon_c/\omega_c)]$ is the Morel-Anderson Coulomb pseudopotential with $\mu = N(0)\langle\langle\tilde{V}(\mathbf{k}-\mathbf{k}')\rangle\rangle_{\rm FS}$ the Coulomb parameter averaged over the Fermi surface (FS), $\varepsilon_c$ and $\omega_c$ denoting characteristic energy scales of the electron and phonon systems, respectively [29, 31, 32]. The highest temperature that Eqs. (14) admit nontrivial solutions $\Delta(i\omega_n) \neq 0$ is defined as $T_c$ [21, 31].

## III. Analytical solution based on the Debye model

One of the widely used phonon models to describe vibrational properties of crystalline solids is the Debye model, in which the Debye frequency ($\omega_{\rm D}$) is introduced as a measure of the maximum phonon energy [33]. The Eliashberg spectral function within the Debye model is given by [34, 35],

$$\alpha^2F(\omega) = \lambda_0 \frac{\omega^2}{\omega_{\rm D}^2}\Theta(\omega_{\rm D}-\omega), \tag{15}$$

where $\lambda_0 = \lambda(0)$ is the electron-phonon coupling constant evaluated at $\omega_n = \omega_{n'}$ in Eq. (13), $\omega_{\rm D}$ denotes the Debye frequency, and $\Theta(x)$ is the Heaviside step function. The Debye frequency $\omega_{\rm D}$ associated with the spectral function can be determined from Eq. (15) as

$$\int_0^\infty \alpha^2F(\omega)\,d\omega = \int_0^\infty \lambda_0 \frac{\omega^2}{\omega_{\rm D}^2}\Theta(\omega_{\rm D}-\omega)\,d\omega = \lambda_0\frac{\omega_{\rm D}}{3}, \tag{16}$$

that is,

$$\omega_{\rm D} = \frac{3}{\lambda_0}\int_0^\infty \alpha^2F(\omega)\,d\omega\,, \tag{17}$$

which contains the information of electron-phonon coupling, in contrast to the standard Debye frequency only representing the vibrational modes.

We now introduce an analytical ansatz for $\Delta(i\bar{\omega}_n)$, inspired by Ref. [20], given by

$$\Delta(i\bar{\omega}_n) = \Delta(i\bar{\omega}_0)\big/(1+\bar{\omega}_n^2\big/\Omega^2)\,, \tag{18}$$

where $\bar{\omega}_n \equiv \omega_n/\omega_{\rm D}$ and $\Omega^2 = \lambda_0/\lambda_\Delta$. The parameter $\lambda_\Delta = 0.7$ defines a characteristic coupling scale that controls the decay of the superconducting gap as a function of Matsubara frequency.

Using Eqs. (15) and (18), we derive in Appendix A the following analytical solutions given by

$$\bar{T}_c^{\rm weak}(\lambda_0,\mu^*) \equiv \frac{T_c^{\rm weak}}{\omega_{\rm D}} \approx \frac{1.13}{\sqrt{\lambda_\Delta/\lambda_0}}\exp\left\{-\frac{1+\lambda_0}{\lambda_0-\mu^*} - \lambda_0\frac{\ln(2)-\ln(\lambda_\Delta/\lambda_0)}{(\lambda_0-\mu^*)(2-\lambda_\Delta/\lambda_0)}\right\} \tag{19}$$

for the weak-coupling regime and

$$\bar{T}_c^{\rm strong} \equiv \frac{T_c^{\rm strong}}{\omega_{\rm D}} \approx \frac{1}{\pi\sqrt{8}}\left(\frac{4\lambda_0}{3+7\mu^*}-1\right)^{1/2} \tag{20}$$

for the ultra-strong-coupling regime. A general analytical solution of the isotropic Eliashberg equations can be obtained by smoothly interpolating between the asymptotic results in Eqs. (19) and (20), yielding

$$\bar{T}_c^{pred}(\lambda_0,\mu^*) \equiv \frac{T_c^{pred}}{\omega_{\rm D}} = \frac{[4\lambda_0/(3+7\mu^*)-1]\Theta[4\lambda_0/(3+7\mu^*)-1]\bar{T}_c^{\rm strong}(\lambda_0,\mu^*) + \lambda_\Delta\bar{T}_c^{\rm weak}(\lambda_0,\mu^*)}{[4\lambda_0/(3+7\mu^*)-1]\Theta[4\lambda_0/(3+7\mu^*)-1]+\lambda_\Delta}. \tag{21}$$

The analytical solution of Eq. (21) is plotted as a red grid in Figure 1 and compared with the self-consistent solution $T_c^{\mathrm{Num}}$ (cyan spheres) obtained from the isotropic Eliashberg equations (14) by using the Debye spectral function in Eq. (15), the condition $\Delta(i\overline{\omega}_n)=0$ for all $n\in\mathbb{Z}$, and 600 Matsubara frequencies in Eq. (14), without the approximation expressed in Eq. (A1).

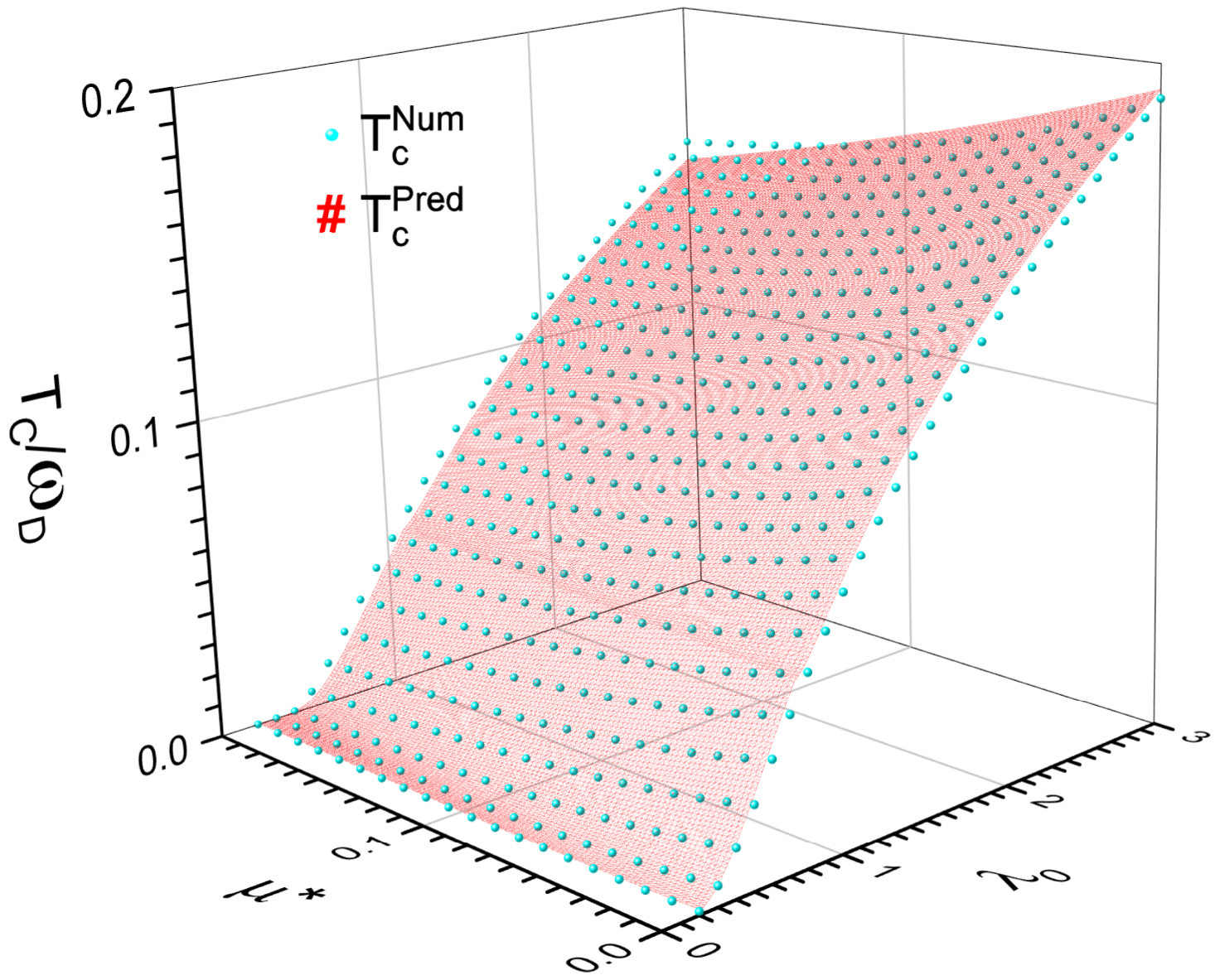

**Figure 1**. Self-consistently calculated critical temperature ($T_c^{\mathrm{Num}}$) from the isotropic Eliashberg equations (14) compared with the analytical prediction ($T_c^{\mathrm{Pred}}$) obtained from Eq. (21), both as functions of the electron-phonon coupling constant ($\lambda_0$) and the Morel-Anderson Coulomb pseudopotential ($\mu^*$).

Observe in Figure 1 the close agreement, even in the strong coupling regime with $\lambda_0\approx 2.5$, between the analytical prediction $T_c^{\mathrm{Pred}}(\lambda_0,\mu^*)$ and the numerically computed $T_c^{\mathrm{Num}}(\lambda_0,\mu^*)$ over a grid of $28\times 20$ points. The mean relative error (MRE) of Figure 1 is $\varepsilon_{\mathrm{MRE}}=6.557\%$.

In the next section, we apply the analytical solution given in Eq. (21) to the strong-coupling hydride superconductor $YH_6$.

## IV. Application to hydride superconductors

In contrast to most other hydride superconductors, which typically require pressures exceeding 200 GPa to achieve a superconducting state [36], the metal hydride $YH_6$ exhibits a remarkably high critical temperature of 224 K at a moderate pressure of just 166 GPa [11].

First-principles calculations of the Eliashberg spectral function [$\alpha^2F(\omega)$] in $YH_6$ are divided into two main stages: (i) Geometry optimization, based on electronic calculations within the density functional theory (DFT); and (ii) Spectral function calculations, carried out using the density functional perturbation theory (DFPT). Both stages are performed by means of the Quantum ESPRESSO code (QE) [37], where the exchange-correlation effects are described by using the Perdew-Burke-Ernzerhof (PBE) functional [38], while the pseudopotentials are constructed using the projector augmented-wave (PAW) method [39]. A summary of computational parameters used in this calculation is provided in Table 1.

**Table 1.** Main parameters used in *ab initio* calculations via the Quantum ESPRESSO code

| Physical quantity | Geometry optimization (DFT) | Spectral function $\alpha^2F(\omega)$ (DFPT) |
|---|---|---|
| Total energy convergence tolerance | $10^{-14}$ Ry | $10^{-15}$ Ry |
| Maximum force convergence tolerance | $10^{-3}$ Ry/Bohr | |
| Maximum stress convergence tolerance | 0.05 kbar | |
| Plane-wave energy cutoff | 100 Ry | 77 Ry |
| Broadening method | Marzari-Vanderbilt | Marzari-Vanderbilt |
| Phonon broadening $\eta_{\text{ph}}$ | | 0.2 THz |
| Electron-phonon interaction broadening $\eta_{\text{el-ph}}$ | | 0.003-0.045 Ry |
| **k**-point grid for electrons | 48×48×48 | 72×72×72 |
| **q**-point grid for phonons | | 9×9×9 |

In Figure 2, using the DFPT parameters of Table 1, the Eliashberg spectral functions $\alpha^2F(\omega)$ of hydride $YH_6$ at 166 GPa possessing a cubic *Im-3m* structure (illustrated in the inset) are shown for two electron-phonon broadening parameters: $\eta_{\text{el-ph}} = 0.03\,\text{Ry}$ (red circles) and $\eta_{\text{el-ph}} = 0.003\,\text{Ry}$ (violet squares).

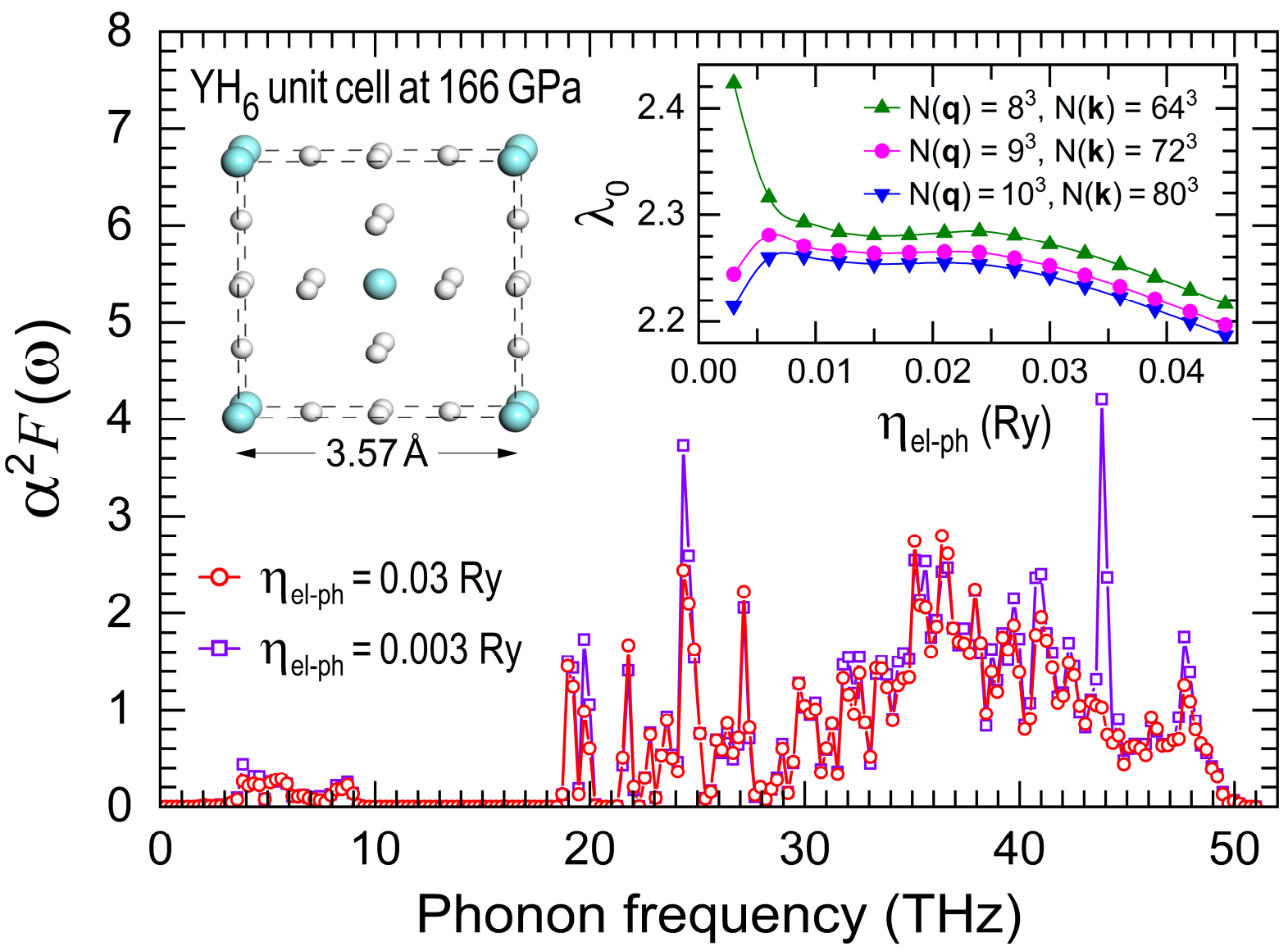

**Figure 2**. Eliashberg spectral functions [ $\alpha^2F(\omega)$ ] with $\eta_{\text{el-ph}} = 0.03\,\text{Ry}$ (open red circles) and $\eta_{\text{el-ph}} = 0.003\,\text{Ry}$ (open violet squares) for $YH_6$, whose yttrium (light blue spheres) and hydrogen (white spheres) atoms are arranged in the cubic *Im-3m* crystal structure. **Inset**: Dependence of $\lambda_0$ on $\eta_{\text{el-ph}}$ for **q**-meshes of 8×8×8 (solid green upward-triangles), 9×9×9 (solid magenta circles), and 10×10×10 (solid blue downward-triangles).

Note in Figure 2 that the sharp peaks become more pronounced as $\eta_{\text{el-ph}}$ decreases, indicating that significantly denser **k-** and **q**-point grids are required for accurate electron and phonon calculations. This behavior is also evident in the inset, where the electron-phonon coupling constant $\lambda_0$ obtained from Eq. (13) is plotted as a function of $\eta_{\text{el-ph}}$ for $\alpha^2F(\omega)$ calculated using three sets of phonon **q**-point grids (with corresponding electron **k**-point grids) of 8×8×8 (64×64×64), 9×9×9 (72×72×72), and 10×10×10 (80×80×80). Observe that the overestimation of $\lambda_0$ in the *ab-initio* calculations diminishes as the number of **q**-points N(**q**) grows. The abrupt increase in $\lambda_0$ observed at $\eta_{\text{el-ph}} = 0.003$ Ry for $N(\mathbf{q}) = 8^3$ is attributed to the insufficient density of **k**- and **q**-grid points in the calculation of $\lambda_0$ for such a small value of $\eta_{\text{el-ph}}$.

The superconducting critical temperature ($T_c$) of $YH_6$ can be self-consistently calculated by solving the isotropic Eliashberg equations (14) using $\alpha^2F(\omega)$ obtained from the *ab initio* calculations, enforcing the condition $\Delta(i\overline{\omega}_n) = 0$ for all $n \in \mathbb{Z}$, as illustrated in Figure A2. Alternatively, $T_c$ of hydride $YH_6$ can also be calculated by using the analytical solution given in Eq. (21), where the numerically obtained spectral function $\alpha^2F(\omega)$ is used to evaluate $\lambda_0(\eta_{\text{el-ph}})$ via Eq. (13) and $\omega_D$ via Eq. (17). For the hydride $YH_6$, a relatively high value of the Morel-Anderson Coulomb pseudopotential $\mu^* = 0.18$ is commonly used [11].

The McMillan-Allen-Dynes formula for the calculation of $T_c$ has a general expression given by [17]

$$T_c = f_1\, f_2\, \frac{\omega_{\ln}}{1.2} \exp\left[ -\frac{1.04(1+\lambda_0)}{\lambda_0 - \mu^* - 0.62\lambda_0\mu^*} \right], \tag{22}$$

where $f_1$ and $f_2$ can be written as

$$f_1 = \left\{ 1 + \left[ \frac{\lambda_0}{2.46\,(1+3.8\mu^*)} \right]^{3/2} \right\}^{1/3} \quad \text{and} \quad f_2 = 1 + \frac{(\omega_2/\omega_{\ln} - 1)\lambda_0^2}{\lambda_0^2 + [1.82(1+6.3\mu^*)\,\omega_2/\omega_{\ln}]^2} \tag{23}$$

with the logarithmic and second spectral moments defined by

$$\omega_{\ln} \equiv \exp\left[ \frac{2}{\lambda_0} \int_0^\infty \frac{d\omega}{\omega} \alpha^2F(\omega) \ln\omega \right] \quad \text{and} \quad \omega_2 \equiv \left\{ \left\langle \omega^2 \right\rangle \right\}^{1/2} = \left\{ \frac{2}{\lambda_0} \int_0^\infty \frac{d\omega}{\omega} \alpha^2F(\omega)\,\omega^2 \right\}^{1/2}. \tag{24}$$

In the weak-coupling regime, the Allen-Dynes correction factors $f_1$ and $f_2$ smoothly approach unity. In this limit, equation (22) reduces to the McMillan form [16]. In addition, Eq. (15) yields $\omega_{\ln} = \omega_D / e^{1/2}$ and $\omega_2 = \omega_D / 2^{1/2}$. These relations show that $\omega_D$ represents the upper cutoff of the Eliashberg spectral function $\alpha^2F(\omega)$, while $\omega_{\ln}$ and $\omega_2$ correspond to different weighted averages of the same spectrum.

Given that the Eliashberg spectral function $\alpha^2F(\omega)$ depends on the electron-phonon broadening parameter ($\eta_{\text{el-ph}}$), the electron-phonon coupling constant ($\lambda_0$) and Debye frequency ($\omega_D$) calculated from Eqs. (13) and (17) have variations $\lambda_0 \in (2.07, 2.4)$ and $\omega_D \in (2084\,\text{K}, 2102\,\text{K})$ for $\eta_{\text{el-ph}} \in (0.003, 0.045)$. In Figure 3, we show the critical temperature $T_c$ as a function of $\eta_{\text{el-ph}}$, obtained from the analytical solution of Eq. (21) (red circles) and from the self-consistent calculation by numerically solving Eqs. (14) based on $\alpha^2F(\omega)$ of $YH_6$ shown in Figure 2 (blue triangles). These results of $T_c$ are additionally compared with those obtained from the McMillan-Allen-Dynes formula given by Eq. (22) (dark yellow rhombuses) [20]. All these theoretical predictions are contrasted to the experimental data (magenta dashed line) reported in Ref. [11]. Notice that the McMillan-Allen-Dynes formula tends to underestimate the superconducting critical temperature. In contrast, the analytical solution based on the Debye model, as well as the self-consistent numerical solution, typically overestimate $T_c$.

The accuracy of the analytical solution (21) can be further validated by applying it to a wide range of superconducting hydrides and comparing the predicted critical temperature, $T_c^{\text{Pred}}$, with the numerical values, $T_c^{\text{Num}}$, obtained from self-consistent solution of the Eliashberg equations (14), using the same Eliashberg spectral function $\alpha^2F(\omega)$ reported in the literature for each compound.

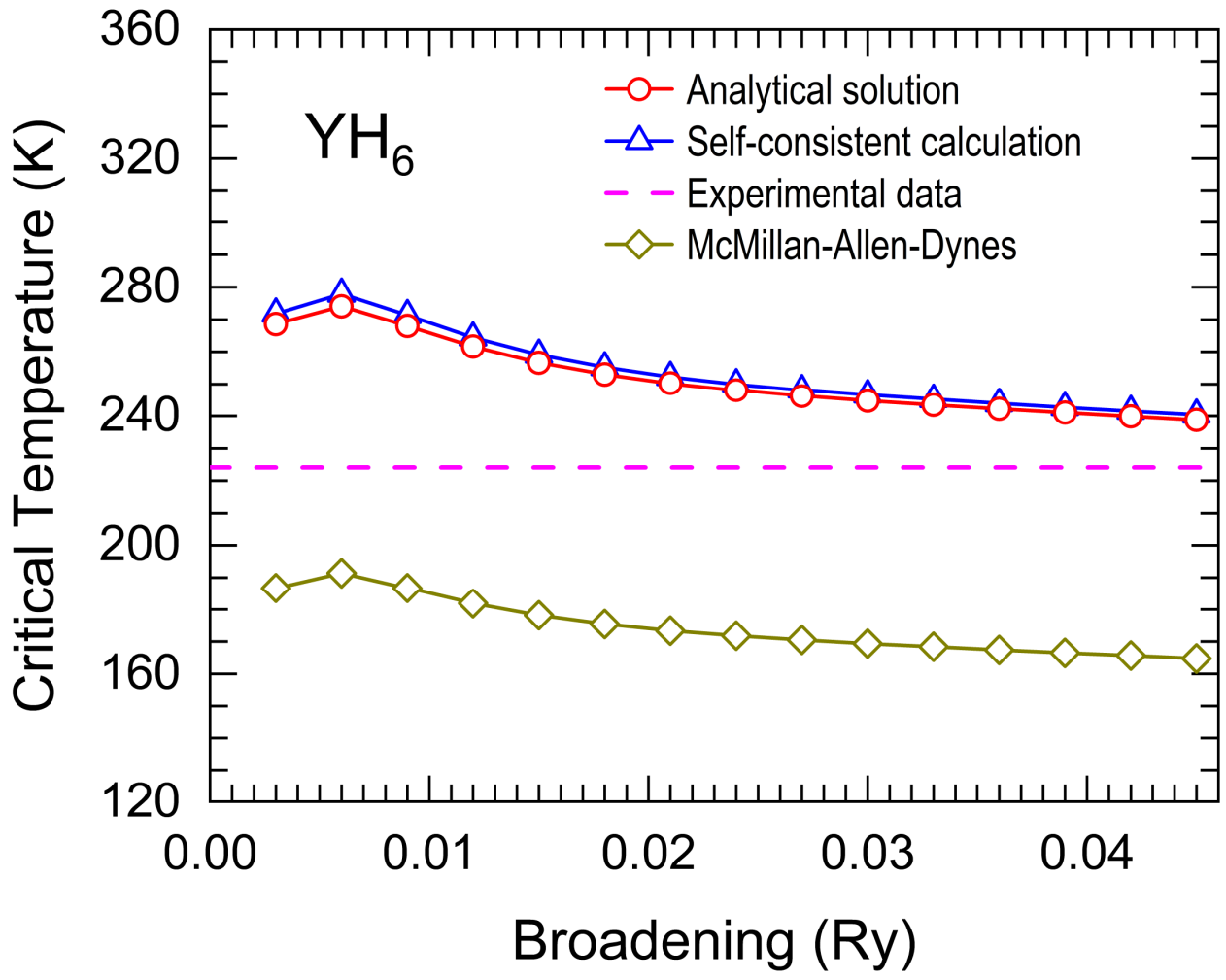

**Figure 3**. Superconducting critical temperatures ($T_c$) as a function of the broadening parameter ($\eta_{\text{el-ph}}$), obtained from the analytical solution in Eq. (21) (red circles) and the self-consistent calculation (blue triangles), are compared with experimental data of $YH_6$ from Ref. [11] (magenta dashed line) and the McMillan-Allen-Dyne formula (dark yellow rhombuses).

A comparison between analytical and numerical critical temperatures, evaluated around the equality line as in Ref. [40], is presented in Figure 4 for the following compounds: $La_3ThH_{40}$ (blue symbols) [41], $ScH_{13}$ (red symbols) [42], $LaThH_{20}$ (green symbols) [41], $LaTh_3H_{40}$ (magenta symbols) [41], $ScCaH_{12}$ (orange symbols) [43], $ThH_9$ (dark cyan symbols) [44], $H_3SXe$ (dark yellow symbols) [45], and $ScCaH_8$ (violet symbols) [43]. This approach yields a mean relative error (MRE) of 1.02% for $\mu^* = 0.15$, improving upon the 5.292% reported in Ref. [40].

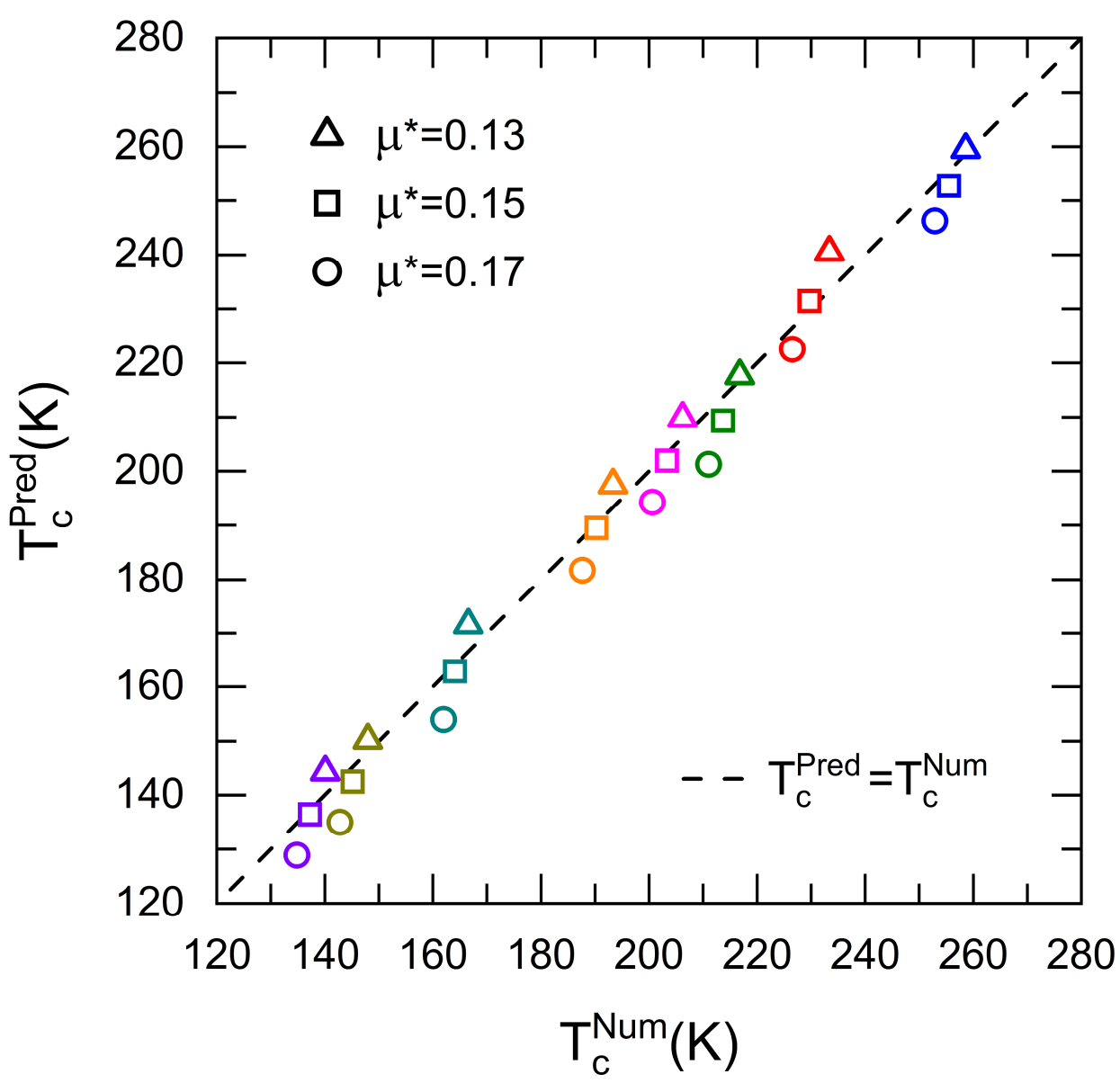

**Figure 4.** Predicted critical temperature $T_c$ as a function of the electron-phonon coupling constant $\lambda_0$, obtained from Eq. (21) with $\mu^* = 0.15$ (red circles). The shaded region is bounded by $\mu^* = 0$ (upper border) and $\mu^* = 0.35$ (lower border). Green triangles: *ab initio* data from Refs. [40-42]. Blue squares: experimental $T_c$ from Ref. [42].

To assess the robustness and scaling behavior of our wide-range analytical solution [see Eq. (21)], Figure 5 compares the predicted critical temperature $T_c$ (red circles) with data from 64 superconducting hydrides at various pressures [46-48]. The predicted values are computed using $\omega_D = 1800\,\mathrm{K}$, $\mu^* = 0.15$, and the Debye spectral function given in Eq. (15), as a function of the electron-phonon coupling constant $\lambda_0$. In Figure 5, all the eleven experimental $T_c$ values (blue squares) are taken from Ref. [48], and for eight of them, the corresponding $\lambda_0$ values were obtained from reported *ab-initio* calculations performed at pressures close to the experimental conditions [49-56]. The light-magenta shaded region in Figure 5 illustrates the variation of the predicted $T_c$ when $\mu^* = 0$ (upper border) and $\mu^* = 0.35$ (lower border) for a given $\lambda_0$.

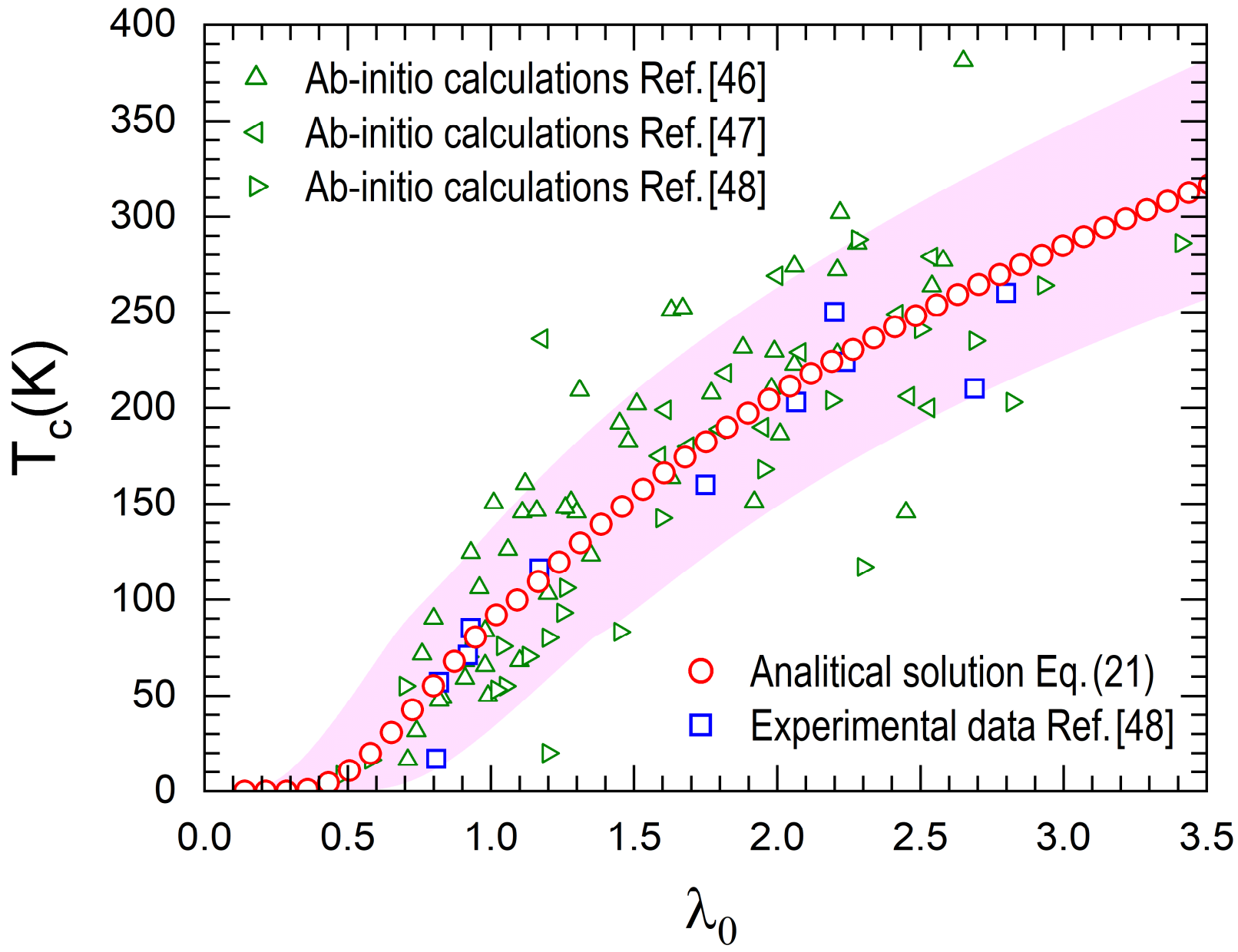

**Figure 5.** Predicted critical temperature $T_c$ as a function of the electron-phonon coupling constant $\lambda_0$, obtained from Eq. (21) with $\mu^* = 0.15$ (red circles). The shaded region is bounded by $\mu^* = 0$ (upper border) and $\mu^* = 0.35$ (lower border). Green triangles: *ab initio* data from Refs. [46-48]. Blue squares: experimental $T_c$ from Ref. [48].

As shown in Figure 5, superconductivity emerges only when the electron-phonon coupling exceeds the Morel-Anderson Coulomb pseudopotential, *i.e.* $\lambda_0 > \mu^*$. The predicted exponential dependence of $T_c(\lambda_0)$ in the weak coupling regime and its transition to a square-root law in the ultra-strong coupling regime are corroborated by statistics from *ab initio* datasets and experiments. In addition, the predicted shaded region of $T_c$ for $0 \le \mu^* \le 0.35$ captures the overall scaling behavior of the ab initio and experimental $T_c$ values.

In general, the superconducting coherence length in the Einstein model for $T \lesssim T_c$ can be written in atomic units as [57] $\xi(T) = v_F / (\lambda_0^{3/2} \omega_E) = 3v_F / (2\lambda_0^{3/2} \omega_D)$, and from the Ginzburg-Landau theory [58] we have $\xi(0) = \xi(T)\sqrt{1 - T/T_c}$. For the case of $YH_6$ at 166 GPa and $T = 0.9T_c$, the Fermi velocity is $v_F = 8.76 \times 10^5\,\mathrm{m/s}$ [59], the electron-phonon coupling constant stands $\lambda_0 = 2.24$ [11], and the Debye frequency obtained from Eq. (17) holds $\omega_D \approx 2100$ K $\approx 43.76$ THz . In consequence, the zero-temperature coherence length is $\xi(0) \approx 1.42$ nm , in close agreement with the experimental data of $\xi(0) \approx 1.45$ nm obtained from $H_{C2}$ measurements [60]. This analysis aligns with the Eliashberg picture, where the finite phonon lifetime $(\tau_{ph} \sim \omega_D^{-1})$ and retardation effects limit the temporal range of the electron-phonon interaction, setting a short pairing length scale comparable to the zero-temperature coherence length $\xi(0) \approx 1.4$ nm .

Moreover, in the Einstein phonon model, the Eliashberg spectral function is [16] $\alpha^2 F(\omega) = (\lambda_0 \omega_E / 2) \delta(\omega - \omega_E)$ , which leads to $\omega_E = (2/\lambda_0) \int_0^\infty \alpha^2 F(\omega) d\omega$ and comparing with Eq. (17) one gets $\omega_E = 2\omega_D/3$ . In consequence, from Eq. (13) we have

$$\lambda(i\omega_n) = \int_0^\infty \frac{2\omega\alpha^2 F(\omega)}{\omega^2 + \omega_n^2} d\omega = \lambda_0 \omega_E \int_0^\infty \frac{\omega\delta(\omega - \omega_E)}{\omega^2 + \omega_n^2} d\omega = \lambda_0 \frac{\omega_E^2}{\omega_E^2 + \omega_n^2} = \lambda_0 \frac{\omega_D^2}{\omega_D^2 + \frac{9}{4}\omega_n^2} . \quad (25)$$

In contrast, Eq. (A1) of the Debye model can be rewritten as $\lambda(i\omega_n) \cong \lambda_0 \omega_D^2 / (\omega_D^2 + 2\omega_n^2)$ , which produces a slower decay of the electron-phonon coupling $\lambda(i\omega_n)$ with the Matsubara frequency $\omega_n$ than Eq. (25), *i.e.*, the Debye model predicts a stronger electron-phonon coupling than that from the Einstein model, for an arbitrary Eliashberg spectral function.

Recent studies have shown that phonon anharmonicity and damping can substantially influence the superconducting critical temperature [61-63]. Likewise, the role of soft phonons and Kohn-like anomalies in enhancing electron-phonon coupling and superconductivity has been discussed by Jiang *et al.* [64,65]. These results suggest that phonon linewidths, commonly represented through a broadening parameter in numerical calculations of $\alpha^2 F(\omega)$ , may have physical implications beyond their computational role [66].

## V. Conclusions

Accurately solving the Eliashberg equations, formulated over sixty-five years ago, remains a notable challenge for theoretical physicists. The McMillan-Allen-Dynes formula, validated across a wide range of superconducting materials, represents a benchmark for new solutions. The recent discovery of hydride superconductors has renewed interest in the phonon-mediated pairing framework described by Eliashberg theory, which has been integrated into the exchange-correlation functionals within the DFT framework [67]. Recently, numerous studies based on the Einstein phonon model have also been carried out [19-24]. However, a more realistic representation of vibrational modes is provided by the Debye model, whose inclusion in the analytical solution to the isotropic Eliashberg equations leads to mathematical complications, and only attempts with coarse approximations have been reported to date [25].

In this article, we have presented an accurate analytical solution to the isotropic Eliashberg equations based on the Debye model, whose accuracy was verified by comparing with the fully self-consistent numerical solutions derived from the Debye spectral function, as shown in Figure 1, and it is proven to be well-suited in weak, strong, and ultra-strong coupling regimes (see Figure A4). The deduction of this solution, inspired by Ref. [20] and discussed in Appendix A, is almost straightforward compared with other analytical derivations. This analytical solution is also verified in the study of a representative strong-coupling hydride superconductor $YH_6$ with an electron-phonon coupling constant of $\lambda_0 \approx 2.2$ at a relatively low pressure of 166 GPa. Notably, the lattice constant of 3.57 Å obtained from DFT calculations agrees well with the experimental value. In addition, the Debye frequency used in the analytical solution, extracted from the spectral function of $YH_6$ via Eq. (17), leads to predictions conducting an exceptional agreement with the self-consistent solution, as observed in Figure 3. It is well known that fully self-consistent solutions of the isotropic Eliashberg equations with a constant $N(0)$ [Eq. (14)] frequently overestimate the superconducting critical temperature [68]. However, using a commonly adopted value of $\eta_{\text{el-ph}} \approx 0.02\,\text{Ry}$ [69,70], the present wide-range analytical solution predicts a critical temperature for $YH_6$ with a relative deviation of only $|T_c^{\text{pred}} - T_c^{\text{exp}}| / T_c^{\text{exp}} \approx 11.4\%$, compared with $22.5\%$ obtained from the McMillan-Allen-Dynes formula, as shown in Figure 3.

This comparison is further extended to eight additional hydride compounds using their respective ab initio spectral functions (see Figure 4), where a strong correlation between the self-consistent and analytical solutions of the Eliashberg equations is clearly observed. Moreover, our wide-coupling-range analytical solution exhibits an exponential dependence of $T_c(\lambda_0)$ in the weak coupling regime and a square-root law in the ultra-strong coupling regime. This fact has been corroborated by many ab initio and experimental $T_c$ results (see Figure 5).

In general, this analytical solution provides a reliable first estimate of the superconducting critical temperature from the Debye phonon frequency, and it can be used as useful starting point for the self-consistent iterations to reduce the computational cost. Moreover, this solution could also be used to estimate the upper bound of $T_c$ even for strong coupling materials. This analytical solution offers an explicit functional form that enhances conceptual understanding, which is invaluable for building physical intuition and for pedagogical purposes. Furthermore, in the context of high-throughput computational searches for new superconductors, a fast and accurate analytical estimate is a powerful tool for pre-screening candidates before committing to more expensive numerical calculations. The study presented in this article can be extended to derive analytical solutions of the non-isotropic Eliashberg equations given by Eqs. (10) to address other boson-mediated anisotropic high-$T_c$ superconductors [71], as well as superconducting nanostructures [72].

Finally, the analytical derivation in the present article is based on the harmonic Debye approximation and therefore does not explicitly include phonon self-energy corrections arising from anharmonicity. A more general treatment would require incorporating frequency-dependent phonon linewidths and renormalized phonon propagators within the Eliashberg formalism. Extending the present analytical framework to account for these effects constitutes an interesting direction for future work.

## Acknowledgments

This work has been supported by the Secretaría de Ciencias, Humanidades, Tecnología e Innovación (SECIHTI) of Mexico under grant CF-2023-I-830 and by the Universidad Nacional Autónoma de México via project PAPIIT-IN116126. Computations were performed at Miztli through grant LANCAD-UNAM-DGTIC-039. The technical assistance of Alejandro Pompa, Oscar Luna, Cain Gonzalez, Silvia E. Frausto, and Yolanda Flores is fully appreciated. T.J.E. acknowledges the master's fellowship from SECIHTI.

## Appendix A. Debye-driven analytical solution of the isotropic Eliashberg equations

The electron-phonon coupling constant $\lambda(i\bar{\omega}_n - i\bar{\omega}_{n'})$, as defined in Eq. (13) for the Debye model, is

$$\lambda(i\bar{\omega}_n - i\bar{\omega}_{n'}) = \int_0^1 \frac{2\bar{\omega}\alpha^2 F(\bar{\omega})}{(\bar{\omega}_n - \bar{\omega}_{n'})^2 + \bar{\omega}^2} d\bar{\omega} = \lambda_0 - \lambda_0 (\bar{\omega}_n - \bar{\omega}_{n'})^2 \ln\left(1 + \frac{1}{(\bar{\omega}_n - \bar{\omega}_{n'})^2}\right) \cong \frac{\lambda_0}{1 + 2(\bar{\omega}_n - \bar{\omega}_{n'})^2}, \quad \text{(A1)}$$

where $\bar{\omega}_n \equiv \omega_n / \omega_D$ is the normalized Matsubara frequency with respect to the Debye frequency $\omega_D$ and $\bar{\omega} \equiv \omega/\omega_D$. In Eq. (A1), $\alpha^2 F(\bar{\omega}) = \lambda_0 \bar{\omega}^2 \Theta(1-\bar{\omega})$ is the Eliashberg spectral function [34, 35], where $\Theta(x)$ is the Heaviside step function and $\lambda_0 = \lambda(0)$ is the electron-phonon coupling constant for $\omega_n = \omega_{n'}$. In general, $\lambda(i\bar{\omega}_n - i\bar{\omega}_{n'})$ for the Debye model can be approximated by $\lambda_0 / [1 + 2(\bar{\omega}_n - \bar{\omega}_{n'})^2]$, which correctly reproduces at both the high- and low-frequency limits of $\lambda(i\bar{\omega}_n - i\bar{\omega}_{n'})$. To assess the quality of this approximation (A1), we define $\Xi_1(x) \equiv \lambda(i\bar{\omega}_n - i\bar{\omega}_{n'})/\lambda_0 = 1 - x^2 \ln(1 + x^{-2})$ and $\Xi_2(x) = (1 + 2x^2)^{-1}$, where $x = \bar{\omega}_n - \bar{\omega}_{n'} = 2\pi(n - n')T/\omega_D$, and then compare them in Figure A1. Notice that $\Xi_1(0) = \Xi_2(0) = 1$ and $\lim_{x\to\infty} \Xi_1(x) = \lim_{x\to\infty} \Xi_2(x) = 0$, while we observe a relative error of

$$\left| \frac{\Xi_1(x) - \Xi_2(x)}{\Xi_2(x)} \right| \leq 10.4\% \quad \text{for all } x \in \mathbb{R}. \quad \text{(A2)}$$

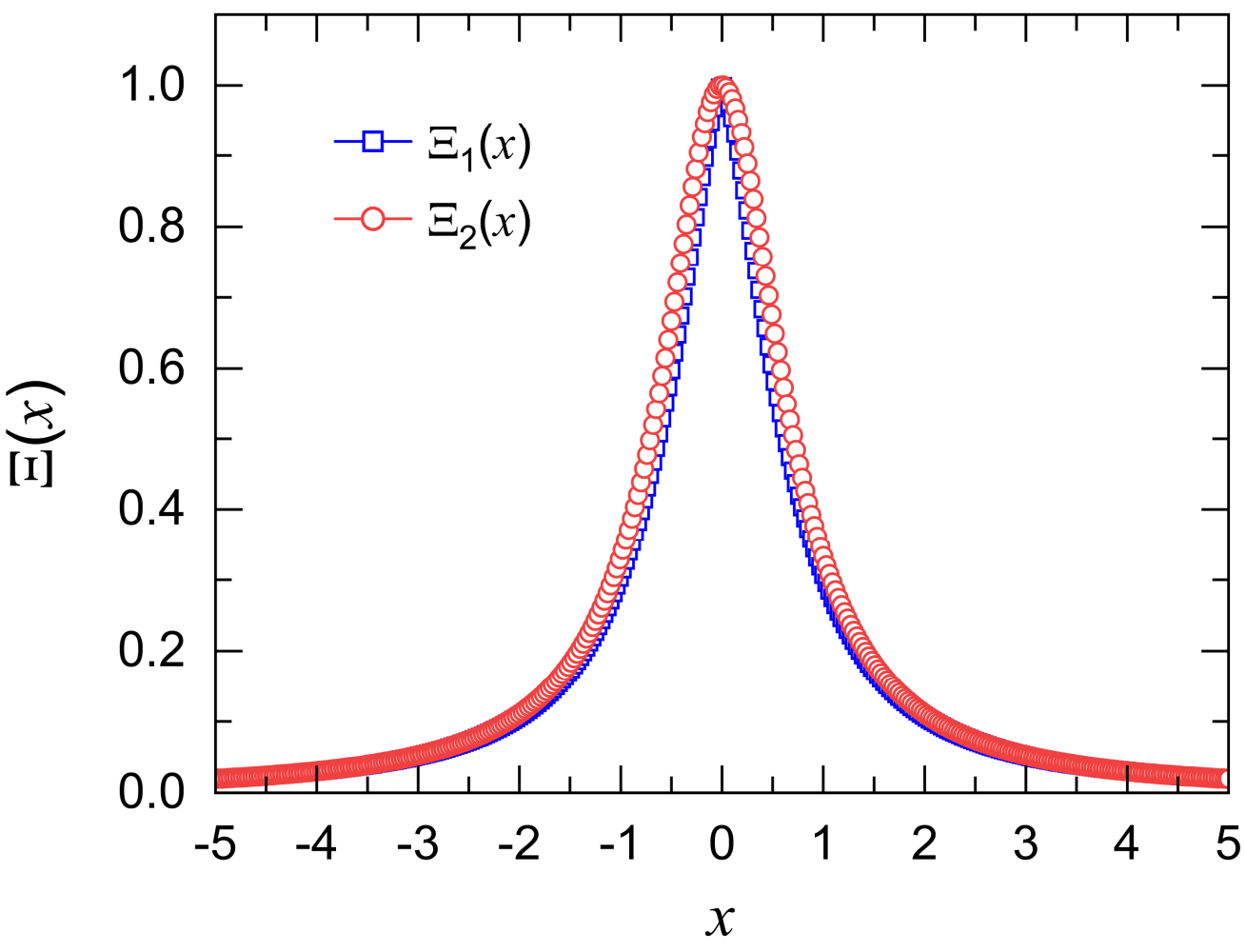

**Figure A1**. Comparison between $\Xi_1(x) = 1 - x^2 \ln(1 + x^{-2})$ (blue squares) and $\Xi_2(x) = (1 + 2x^2)^{-1}$ (red circles) of Eq. (A1), where $x = \bar{\omega}_n - \bar{\omega}_{n'}$.

For temperatures slightly below the critical temperature $T \lesssim T_c$, where the superconducting gap satisfies $\Delta \ll \omega_n$, the isotropic Eliashberg equations (14) can be rewritten as [20]

$$\begin{cases} Z(i\bar{\omega}_n) = 1 + \dfrac{\pi \bar{T}_c}{\bar{\omega}_n} \sum_{n'} \dfrac{\bar{\omega}_{n'}}{|\bar{\omega}_{n'}|} \lambda(i\bar{\omega}_n - i\bar{\omega}_{n'}) \\ \Delta(i\bar{\omega}_n) Z(i\bar{\omega}_n) = \pi \bar{T}_c \sum_{n'} \dfrac{\Delta(i\bar{\omega}_{n'})}{|\bar{\omega}_{n'}|} \left[ \lambda(i\bar{\omega}_n - i\bar{\omega}_{n'}) - \mu^* \right] \end{cases}. \quad \text{(A3)}$$

where $\bar{T}_c \equiv T_c/\omega_{\rm D}$ is the normalized critical temperature and $\bar{T}_c \ll 1$ for most hydride superconductors [73]. In Eq. (A1) $\lambda(\varsigma)$ with $\varsigma \equiv i\bar{\omega}_n - i\bar{\omega}_{n'}$ is an even function, and then the first equation in Eqs. (A3) becomes

$$Z(i\bar{\omega}_0) = 1 + \frac{\pi\bar{T}_c}{\bar{\omega}_0}\left[\lambda(0) + \sum_{n'=1}^{\infty}\lambda(2\pi n'\bar{T}_c) - \sum_{n'=-\infty}^{-1}\lambda(2\pi n'\bar{T}_c)\right] = 1 + \lambda_0, \tag{A4}$$

where $\bar{\omega}_0 = \pi\bar{T}_c$. Substituting Eq. (A4) into the second equation of Eqs. (A3), we obtain for $n = 0$

$$\Delta(i\bar{\omega}_0)(1+\lambda_0) = \pi\bar{T}_c \sum_{n'} \frac{\Delta(i\bar{\omega}_{n'})}{|\bar{\omega}_{n'}|}\left[\lambda(i\bar{\omega}_0 - i\bar{\omega}_{n'}) - \mu^*\right]. \tag{A5}$$

We now introduce an approximate analytical expression for $\Delta(i\bar{\omega}_n)$, inspired by Ref. [16], given by

$$\Delta(i\bar{\omega}_n) = \frac{\Delta(i\bar{\omega}_0)}{1 + \bar{\omega}_n^2/\Omega^2}, \tag{A6}$$

where $\Omega^2 = \lambda_0/\lambda_\Delta$. The parameter $\lambda_\Delta$ defines a characteristic coupling scale associated with the frequency decay of $\Delta(i\bar{\omega}_n)$, quantifying its dynamical stiffness with respect to the Matsubara frequencies. Eq. (A6) is closely analogous to $\Delta(i\bar{\omega}_n) = \Delta(i\bar{\omega}_0)/(1+\omega_n^2/\omega_{\rm E}^2)$ from the Einstein model [21], except for $\bar{\omega}_n^2/\Omega^2 = \omega_n^2/(\omega_{\rm D}^2\Omega^2)$ in Eq. (A6) for the Debye model. From a variational analysis of the half-width at half-maximum of $\Delta(i\bar{\omega}_n)$, or $\Omega^2(\lambda_0)$, to minimize the difference between the numerical and analytical results over the range of $0.1 < \lambda_0 < 3.0$, we find $\lambda_\Delta \approx 0.7$.

The expression (A6) is plotted in Figure A2 for $\lambda_\Delta = 0.7$ and $\lambda_0 = 1, 2$ and $3$, compared to those obtained from the self-consistent solutions of Eqs. (14) using the Eliashberg spectral function of the Debye model given in Eq. (15).

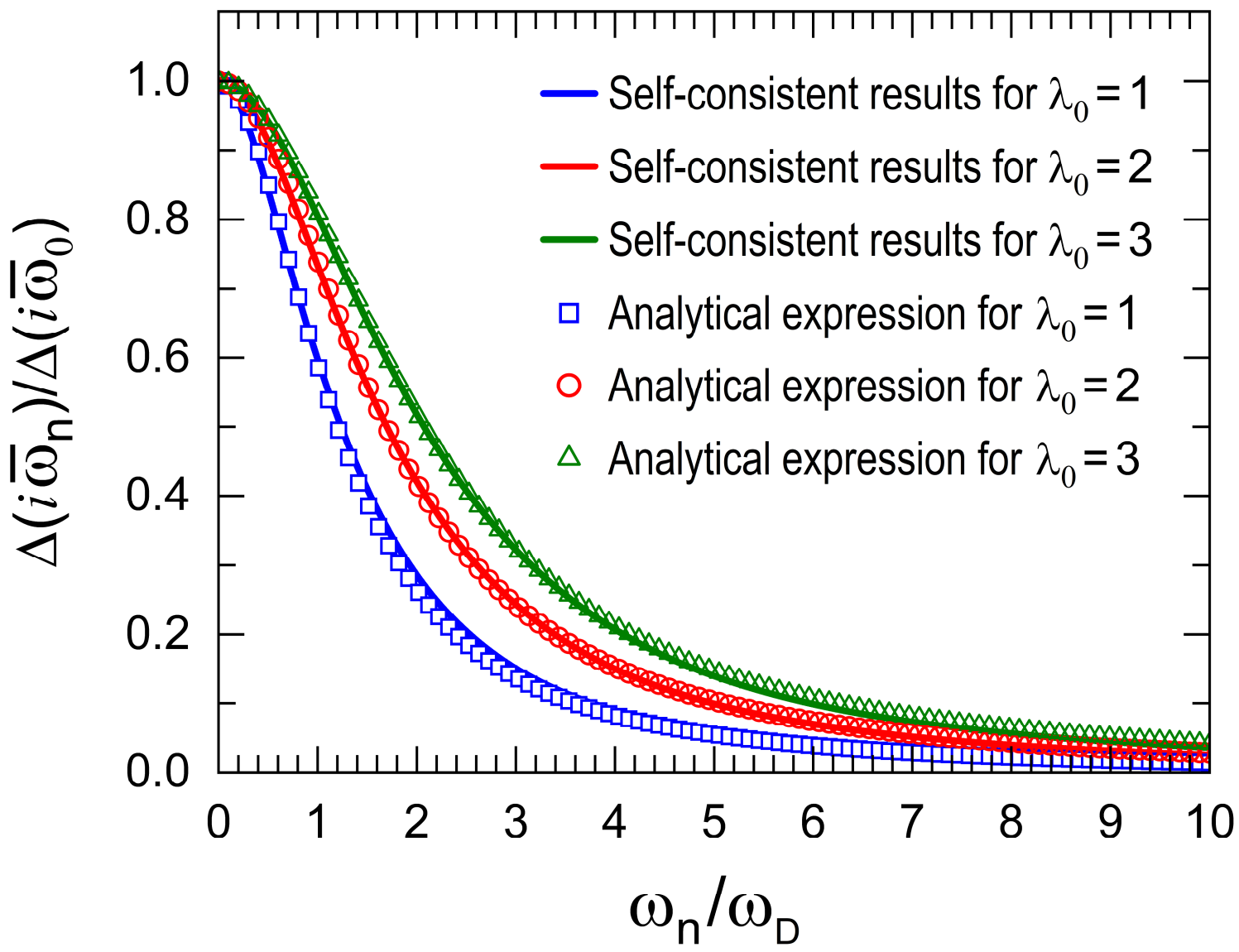

**Figure A2.** Superconducting gap $\Delta(i\omega_n)/\Delta(i\omega_0)$ versus the Matsubara frequencies $\omega_n/\omega_{\rm D}$ self-consistently obtained from the isotropic Eliashberg equations (14) using the Debye-Eliashberg spectral function [Eq. (15)], for the electron-phonon coupling constants $\lambda_0 = 1$ (blue line), $\lambda_0 = 2$ (red line), and $\lambda_0 = 3$ (green line), in comparison with those obtained from Eq. (A6) respectively represented by blue squares, red circles, and green tringles.

Utilizing Eqs. (A1), (A5) and (A6), one gets

$$1+\lambda_0 \cong \pi\bar{T}_c \sum_{n'=-\infty}^{\infty} \frac{1}{|\varpi_{n'}|} \frac{1}{1+\varpi_{n'}^2/\Omega^2} \left[ \frac{\lambda_0}{1+2(\varpi_0-\varpi_{n'})^2} - \mu^* \right], \tag{A7}$$

where

$$\frac{1}{1+2(\varpi_0-\varpi_{n'})^2} = \frac{1}{1+2\varpi_0^2}\left(1+\frac{4\varpi_0\varpi_{n'}-2\varpi_{n'}^2}{1+2(\varpi_0-\varpi_{n'})^2}\right) \simeq \frac{1}{1+2\varpi_0^2}\left(1-\frac{2\varpi_{n'}^2}{1+2\varpi_{n'}^2}\right), \tag{A8}$$

because $\varpi_0 \ll |\varpi_{n'}|$ for most summation terms with $n' \neq 0$ in (A7) and then the approximation given in Eq. (A8) only slightly alters the outcome of Eq. (A7), as discussed in Ref. [20]. Hence, Eq. (A7) becomes

$$\begin{aligned} 1+\lambda_0 &\simeq \pi\bar{T}_c \sum_{n'=-\infty}^{\infty} \frac{1}{|\varpi_{n'}|}\frac{1}{1+\varpi_{n'}^2/\Omega^2}\left(\frac{1}{1+2\varpi_0^2}\lambda_0-\mu^*\right) - \frac{\lambda_0\pi\bar{T}_c}{1+2\varpi_0^2}\sum_{n'=-\infty}^{\infty}\frac{1}{|\varpi_{n'}|}\frac{1}{1+\varpi_{n'}^2/\Omega^2}\frac{2\varpi_{n'}^2}{1+2\varpi_{n'}^2}, \\ &\approx I_0(\xi)\left(\frac{\lambda_0}{1+2\varpi_0^2}-\mu^*\right) - \frac{\lambda_0}{1+2\varpi_0^2}\int_{-\infty}^{\infty}\frac{1}{|x|}\frac{1}{1+x^2/\Omega^2}\frac{x^2}{1+2x^2}dx \end{aligned} \tag{A9}$$

and using the asymptotic expansion of digamma functions ($\psi$) [17, 74], we have

$$I_0(\Omega) \equiv \pi\bar{T}_c \sum_{n=-\infty}^{\infty} \frac{1}{|\varpi_n|(1+\varpi_n^2/\Omega^2)} = \frac{1}{2}\left[\psi\left(\frac{\pi\bar{T}_c\Omega^{-1}-i}{2\pi\bar{T}_c\Omega^{-1}}\right)+\psi\left(\frac{\pi\bar{T}_c\Omega^{-1}+i}{2\pi\bar{T}_c\Omega^{-1}}\right)-2\psi\left(\frac{1}{2}\right)\right] \cong \ln\left(\frac{2e^{\gamma}}{\pi\bar{T}_c\Omega^{-1}}\right). \tag{A10}$$

In Eq. (A9), $\varpi_{n'} = (2n'+1)\pi\bar{T}_c$ and the standard Riemann sum identity is used for $\delta x = 2\pi\bar{T}_c$ given by

$$\int_{-\infty}^{\infty} f(x)dx \approx \sum_{n=-\infty}^{\infty} f(n\,\delta x)\,\delta x\,. \tag{A11}$$

By analytically evaluating the integral in Eq. (A9), we obtain the following equation

$$1+\lambda_0 \approx I_0(\Omega)\left(\frac{\lambda_0}{1+2\varpi_0^2}-\mu^*\right) - \frac{\lambda_0}{1+2\varpi_0^2}\frac{\ln(2)-\ln(\Omega^{-1})}{2-\Omega^{-1}}, \tag{A12}$$

which, using Eq. (A10) and $\varpi_0^2 = \pi^2\bar{T}_c^2 \ll 1$, leads to an analytical weak-coupling solution for $T_c$ given by

$$T_c^{\text{weak}}(\lambda_0,\mu^*) \approx \frac{1.13\omega_{\text{D}}}{\sqrt{\lambda_\Delta/\lambda_0}}\exp\left\{-\frac{1+\lambda_0}{\lambda_0-\mu^*} - \lambda_0\frac{\ln(2)-\ln(\lambda_\Delta/\lambda_0)}{(\lambda_0-\mu^*)(2-\lambda_\Delta/\lambda_0)}\right\}. \tag{A13}$$

For the opposite limit, *i.e.*, the ultra-strong-coupling regime, where $\Delta(i\varpi_n)/\Delta(i\varpi_0)$ is essentially equal to one for $\varpi_0$ and $\varpi_{\pm1}$, as a limiting behavior of Figure A2, the superconducting critical temperature $\bar{T}_c \equiv T_c/\omega_{\text{D}}$ can be calculated from Eq. (A7) by retaining only the three dominant terms with $n'=0$, $n'=-1$ and $n'=1$ in the summation. Consequently, Eq. (A7) can be rewritten as

$$1+\lambda_0 = -\mu^* + \lambda_0 + \frac{\lambda_0}{1+8(\pi\bar{T}_c)^2} - \mu^* + \frac{1}{(3)}\frac{\lambda_0}{1+8(\pi\bar{T}_c)^2} - \frac{1}{3}\mu^*. \tag{A14}$$

Eq. (A14) conduces to an analytical solution for the ultra-strong-coupling regime given by

$$\bar{T}_c^{\text{strong}} \equiv \frac{T_c^{\text{strong}}}{\omega_{\text{D}}} \approx \frac{1}{\pi\sqrt{8}}\left(\frac{4\lambda_0}{3+7\mu^*}-1\right)^{1/2}, \tag{A15}$$

which yields a real number of $T_c$ only for $\lambda_0 \geq (3+7\mu^*)/4$, which is larger than the weak-coupling characteristic value $\lambda_\Delta = 0.7$.

A wide-range analytical solution to the isotropic Eliashberg equations can be built by interpolating between Eqs. (A13) and (A15) yielding [75]

$$\bar{T}_c^{pred}(\lambda_0,\mu^*)=\frac{[4\lambda_0/(3+7\mu^*)-1]\Theta[4\lambda_0/(3+7\mu^*)-1]\bar{T}_c^{\rm strong}(\lambda_0,\mu^*)+\lambda_\Delta\bar{T}_c^{\rm weak}(\lambda_0,\mu^*)}{[4\lambda_0/(3+7\mu^*)-1]\Theta[4\lambda_0/(3+7\mu^*)-1]+\lambda_\Delta},\qquad\text{(A16)}$$

where the weak- and strong-coupling weight factors are $\lambda_\Delta$ and $[4\lambda_0/(3+7\mu^*)-1]\Theta[4\lambda_0/(3+7\mu^*)-1]$, respectively, since the former is the characteristic value $\lambda_0$ in the weak-coupling regime and the latter ensures a smooth entrance of the $\bar{T}_c^{\rm strong}(\lambda_0,\mu^*)$ participation. Equation (A16) constitutes the main analytical result of this article, providing a wide-range estimation of $T_c$ from the Eliashberg spectral function within the Debye model, including the effects of strong electron-phonon coupling and Coulomb pseudopotential corrections.

Alternatively, the critical temperature $T_c$ can also be determined by self-consistently solving the isotropic Eliashberg equations (14) without using the approximation given in Eq. (A1), under the condition $\Delta(i\bar{\omega}_n)=0$ for all $n\in\mathbb{Z}$ when $T=T_c$, as illustrated in Figure A3. In this numerical scheme, $T_c$ is defined as the temperature for which the superconducting gap vanishes at the lowest Matsubara frequency, *i.e.*, $\Delta(i\bar{\omega}_0)=0$ [20,76].

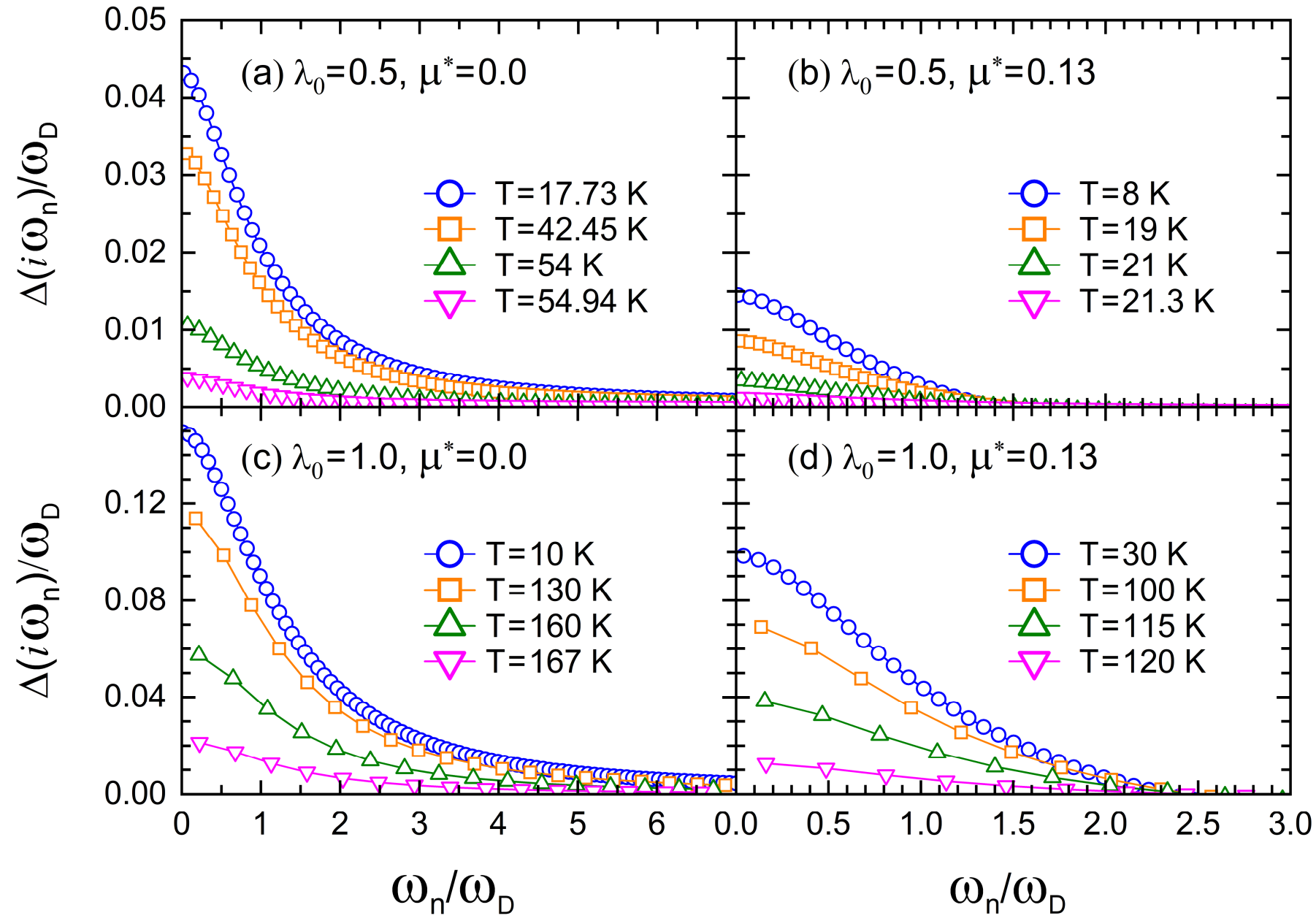

**Figure A3.** Normalized superconducting gap $\Delta(i\omega_n)/\omega_{\rm D}$ obtained from the isotropic Eliashberg equations (14), using the Debye-Eliashberg spectral function [Eq. (15)], for the electron-phonon coupling constants (a,b) $\lambda_0=0.5$ and (c,d) $\lambda_0=1.0$, as well as the Morel-Anderson Coulomb pseudopotentials (a,c) $\mu^*=0.0$ and (b,d) $\mu^*=0.13$.

Using the self-consistent numerical solution ($T_c^{\rm num}$) as a reference, we calculate the mean relative error (MRE) over a two-dimensional grid of $28\times 20$ points in the parameter space defined by $0.2<\lambda_0<3.0$ and $0<\mu^*<0.2$. The resulting MRE is $\varepsilon_{\rm MRE}=6.557\%$, defined as

$$\varepsilon_{\text{MRE}} = \frac{1}{28 \times 20} \sum_{l=1}^{28} \sum_{j=1}^{20} \frac{\left| T_c^{\text{num}}[\lambda_0(l), \mu^*(j)] - T_c^{\text{pred}}[\lambda_0(l), \mu^*(j)] \right|}{\left| T_c^{\text{num}}[\lambda_0(l), \mu^*(j)] \right|}, \tag{A17}$$

where $T_c^{\text{pred}}$ is the analytical prediction given by Eq. (A16). The comparison between $T_c^{\text{num}}(\lambda_0, \mu^*)$ and $T_c^{\text{pred}}(\lambda_0, \mu^*)$ is shown in Figure 1 of the main text.

Finally, the analytical solution of $\bar{T}_c^{pred}$ given in Eq. (A16) is plotted in Figure A4 and compared with the self-consistent results obtained from Eqs. (A3), which are particularly suitable for calculating $T_c$ at $\Delta \to 0$.

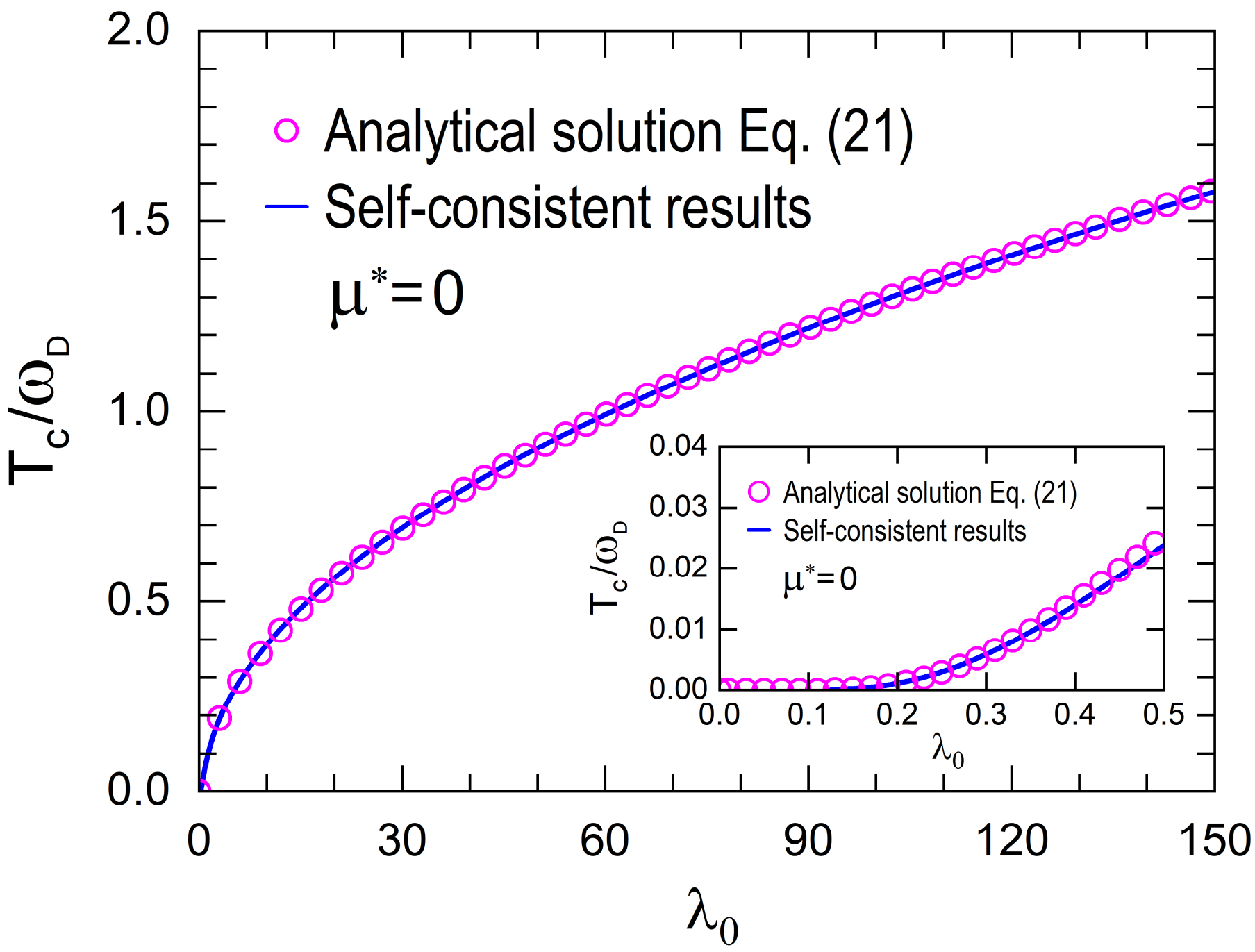

**Figure A4.** Normalized superconducting critical temperature $T_c/\omega_{\text{D}}$ as a function of the electron-phonon coupling parameter $\lambda_0$, obtained from the analytical solution of Eq. (21) (magenta circles), in comparison with the self-consistent results from Eqs. (A3) (blue line), covering from the ultra-strong to weak coupling regimes. The inset presents a magnified view of the weak-coupling region.

Observe in Figure A4 the good agreement between the numerical results and the wide-range analytical solution, including in the weak-coupling regime displayed in the insert. It would be worth noting that the ultra-strong coupling analytical solution expressed in terms of algebraic functions as in Eq. (A15) was highlighted very recently as a significant challenge for the $T_c$ prediction within the Eliashberg superconducting theory [77].